%% file: melvin.tex
\documentclass[aps, prl, twocolumn, reprint,preprintnumbers, nofootinbib, showpacs, superscriptaddress, english]{revtex4}
\usepackage{graphicx, amsmath, amssymb, amsfonts, pifont, bm, cancel}
\usepackage[usenames]{color}
\usepackage{subfigure}
\usepackage{braket}
\usepackage{xspace}
\usepackage{lipsum, babel}
\usepackage[normalem]{ulem} 
\usepackage{xcolor}
\usepackage{pbox}

\newcommand*{\melvin}{{\small M}{\scriptsize ELVIN}\xspace}
\newcommand{\TN}[1]{\textnormal{#1}}

\begin{document}

\widetext
\title{Automated Search for new Quantum Experiments}

\author{Mario Krenn}
\email{mario.krenn@univie.ac.at}
\affiliation{Vienna Center for Quantum Science and Technology (VCQ), Faculty of Physics, University of Vienna, Boltzmanngasse 5, A-1090 Vienna, Austria.}
\affiliation{Institute for Quantum Optics and Quantum Information (IQOQI), Austrian Academy of Sciences, Boltzmanngasse 3, A-1090 Vienna, Austria.}
\author{Mehul Malik}
\affiliation{Vienna Center for Quantum Science and Technology (VCQ), Faculty of Physics, University of Vienna, Boltzmanngasse 5, A-1090 Vienna, Austria.}
\affiliation{Institute for Quantum Optics and Quantum Information (IQOQI), Austrian Academy of Sciences, Boltzmanngasse 3, A-1090 Vienna, Austria.}
\author{Robert Fickler}
\affiliation{Vienna Center for Quantum Science and Technology (VCQ), Faculty of Physics, University of Vienna, Boltzmanngasse 5, A-1090 Vienna, Austria.}
\affiliation{Institute for Quantum Optics and Quantum Information (IQOQI), Austrian Academy of Sciences, Boltzmanngasse 3, A-1090 Vienna, Austria.}
\affiliation{present address: Department of Physics and Max Planck Centre for Extreme and Quantum Photonics, University of Ottawa, Ottawa, K1N 6N5, Canada.}
\author{Radek Lapkiewicz}
\affiliation{Vienna Center for Quantum Science and Technology (VCQ), Faculty of Physics, University of Vienna, Boltzmanngasse 5, A-1090 Vienna, Austria.}
\affiliation{Institute for Quantum Optics and Quantum Information (IQOQI), Austrian Academy of Sciences, Boltzmanngasse 3, A-1090 Vienna, Austria.}
\affiliation{present address: Faculty of Physics, University of Warsaw, Pasteura 5, 02-093 Warsaw, Poland.}
\author{Anton Zeilinger}
\affiliation{Vienna Center for Quantum Science and Technology (VCQ), Faculty of Physics, University of Vienna, Boltzmanngasse 5, A-1090 Vienna, Austria.}
\affiliation{Institute for Quantum Optics and Quantum Information (IQOQI), Austrian Academy of Sciences, Boltzmanngasse 3, A-1090 Vienna, Austria.}

\date{\today}

\begin{abstract}
Quantum mechanics predicts a number of at first sight counterintuitive phenomena. It is therefore a question whether our intuition is the best way to find new experiments. Here we report the development of the computer algorithm \melvin which is able to find new experimental implementations for the creation and manipulation of complex quantum states. And indeed, the discovered experiments extensively use unfamiliar and asymmetric techniques which are challenging to understand intuitively. The results range from the first implementation of a high-dimensional Greenberger-Horne-Zeilinger (GHZ) state, to a vast variety of experiments for asymmetrically entangled quantum states -- a feature that can only exist when both the number of involved parties and dimensions is larger than 2. Additionally, new types of high-dimensional transformations are found that perform cyclic operations. \melvin autonomously learns from solutions for simpler systems, which significantly speeds up the discovery rate of more complex experiments. The ability to automate the design of a quantum experiment can be applied to many quantum systems and allows the physical realization of quantum states previously thought of only on paper.
\end{abstract}

\maketitle

Quantum mechanics encompasses a wide range of counterintuitive phenomena such as teleportation \cite{bennett1993teleporting, wang2015quantum}, quantum interference \cite{hong1987measurement}, quantum erasure \cite{scully1982quantum}, and entanglement \cite{schrodinger1935gegenwartige, einstein1935can, bell1964einstein, greenberger1989going, huber2013structure, lawrence2014rotational}. Despite our struggle to reconcile them with our picture of reality, these phenomena serve as building blocks for many exciting and useful quantum technologies such as quantum cryptography \cite{ekert1991quantum, vazirani2014fully}, computation \cite{shor1994algorithms, rebentrost2014quantum}, and metrology \cite{boto2000quantum, toth2014quantum}. A significant challenge arises, however, when we try to combine such phenomena in order to perform a complex quantum task. Understanding the outcome of even a simple combination of these quantum building blocks can be daunting for the human intuition. Therefore it is natural to ask: Given a certain desired property of a quantum system, what combination of quantum building blocks will be successful in achieving it? 

In order to answer this question, we develop a classical computer algorithm called \melvin, to which we teach how these quantum phenomena work, and subsequently assign it a specific problem. The machine then takes on the task of finding and optimizing arrangements of quantum building blocks that result in a solution. This allows us to uncover experimental methods to create an array of new types of entangled states previously thought to exist only in theory. In addition, it also allows us to address the question of how to manipulate such high-dimensional quantum states, which is key for their use in quantum information systems.

While searching for these experiments, \melvin enlarges its own toolbox by identifying useful groups of elements, leading to a significant speed-up in subsequent discoveries. The experiments found by our algorithm show a departure from conventional experiments in quantum mechanics in that they rely on highly unfamiliar, but perfectly conceivable experimental techniques. This provides some insight into the kind of out-of-the-box thinking that is required for creating such complex quantum states. 

Our method aims to create and manipulate general complex quantum states for which arbitrary transformations are not known. The algorithm creates experiments using experimental accessible optical components that can readily be implemented in the laboratory \cite{malik2015multi, schlederer2015cyclic}. In addition, our algorithm considers multiple degrees of freedom of single quantum systems, and can be extended to include nonlinear components and states more complex than single photons. This would allow us to investigate many other interesting quantum phenomena such as NOON states \cite{afek2010high}, induced coherence \cite{zou1991induced, lemos2014quantum}, quantum teleportation of more complex systems \cite{wang2015quantum} or quantum metrology \cite{boto2000quantum, toth2014quantum}. A complementary field is computer-assisted or automated quantum circuit synthesis (QCS) \cite{shende2006synthesis, maslov2008quantum, saeedi2013synthesis, bocharov2015efficient1, bocharov2015efficient2}, where optimal implementations for quantum algorithms are designed from universal sets of known quantum gates. While very powerful in its own right, the technique of QCS is used for linear qubit networks and usually requires fault-tolerant quantum computers for the implementation of its results.

\begin{figure}[t]
\centerline{\includegraphics[width =0.33 \textwidth]{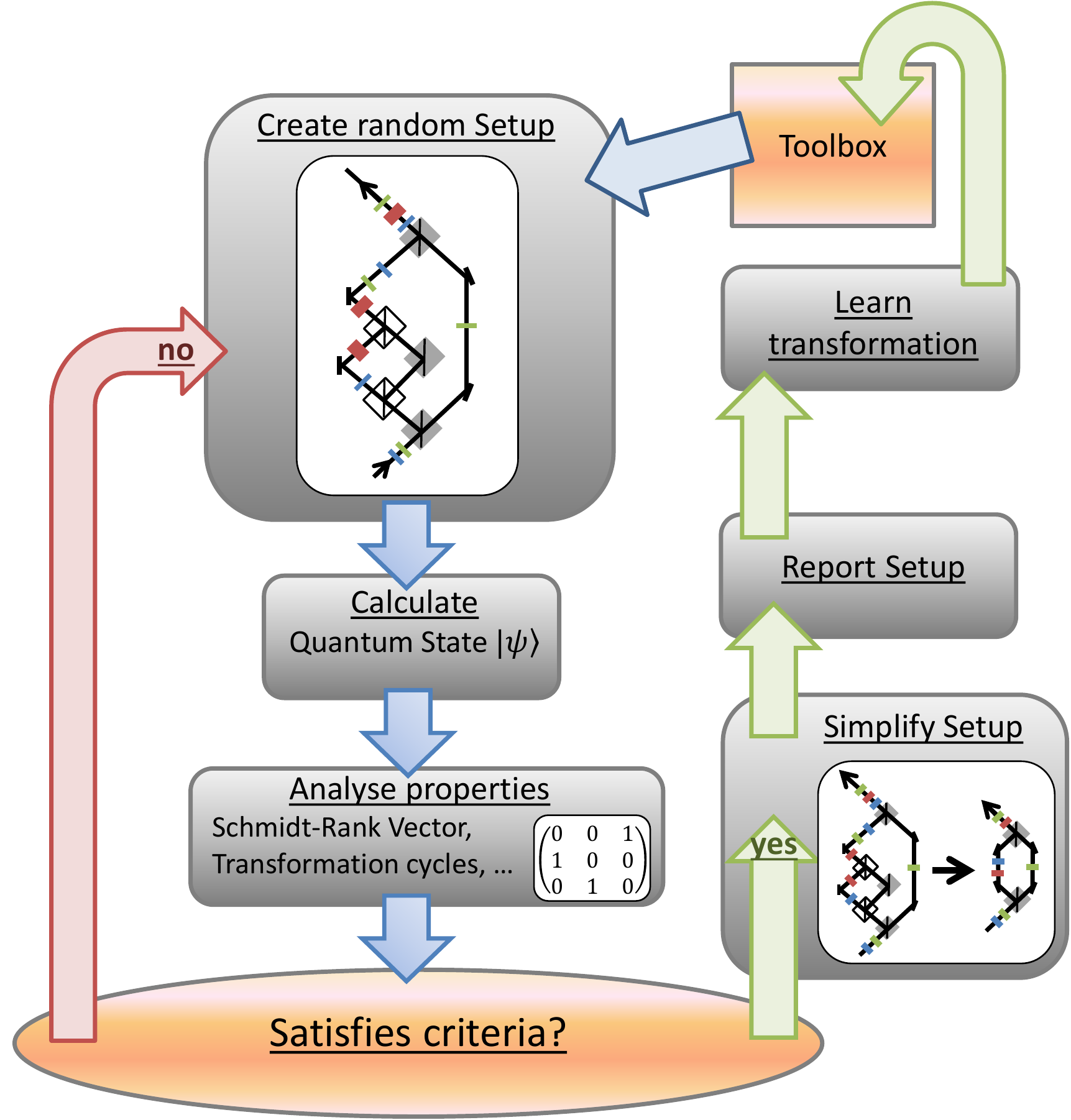}}
\caption{Working principle of the algorithm. First, an experiment is created using elements from a basic toolbox. Then, the quantum state is calculated, and subsequently its properties are analyzed. Those properties are compared with a number of criteria. If these criteria are not satisfied, the algorithm starts over again. However, if the criteria are satisfied, the experiment is simplified and reported, together with all relevant information for the user. Useful solutions can be stored and used in future experiments, which significantly increases the discovery rate of more complex experiments. The orange boxes (toolbox and criteria) are adapted when a different type of quantum property is investigated, while the rest of the algorithm stays the same.}
\label{fig:algorithm}
\end{figure}
\textit{The algorithm} - The main goal is to develop an algorithm which finds experimental implementations for quantum states or quantum transformations with interesting properties, see fig.(\ref{fig:algorithm}). Specific possible input states and a toolbox of experimentally known transformations utilizable by \melvin are defined initially. Using the elements from the toolbox, the algorithm assembles new experiments by arranging elements randomly. Then, from the initial state the resulting quantum state and transformation is calculated and its properties are analyzed. Well-defined criteria that are provided by the user decide whether the calculated quantum state has the desired properties. If the quantum state's properties satisfy the criteria, the experimental configuration is simplified and reported to the user. \melvin can store the configuration in order to use it as a basic building element in subsequent trials. By extending the initial toolbox, it is learning from experience, which leads to a significant speed-up in discoveries of more complex solutions.

All quantum states are calculated using symbolic algebra. Every experimental element is a symbolic modification of the input state. As an example, a 50/50 symmetric non-polarizing beam splitter for photons is described by
\begin{eqnarray}
\textnormal{BS}[\psi,\textnormal{a},\textnormal{b}] = \psi \Leftarrow \left\{
  \begin{array}{lr}
    \textnormal{a}[\ell]\rightarrow \frac{1}{\sqrt{2}}\left(\textnormal{b}[\ell] + i\cdot\textnormal{a}[-\ell]\right)\\
    \textnormal{b}[\ell]\rightarrow \frac{1}{\sqrt{2}}\left(\textnormal{a}\textnormal[\ell] + i\cdot \textnormal{b}[-\ell]\right)
  \end{array}
\right.
\label{eq:symbBS}
\end{eqnarray}
where "$\Leftarrow$" stands for a symbolic replacement followed by a list of substitution rules. $\ell$ stands for the orbital angular momentum (OAM) quantum number of the photon, and a and b denote the input paths of the beam splitter. For simplicity, all other degrees of freedom (such as polarization or frequency) are considered to be the same for all photons. For example, for the two-photon state $\psi=\textnormal{a}[3]\cdot\textnormal{b}[-3]$ (a and b represent the path of one photon, +3 and -3 stand for the OAM of the photon) the beam splitter in path a and b will lead to photon bunching, $\textnormal{BS}[\psi,\textnormal{a},\textnormal{b}] \rightarrow \left(a[-3]^2 + b[3]^2 \right)$, which is the well-known Hong-Ou-Mandel effect \cite{hong1987measurement}. By realizing the calculations with symbolic algebra, adding new elements or even new degrees of freedom is very easy \cite{Supplementary}. Furthermore, it allows easy human-readable intermediate forms, important for the examination of solutions and the novel techniques found by the algorithm.

Next we demonstrate the working principles using two concrete examples. The demonstrations work in the regime of photonic quantum experiments, but the algorithm can readily be adapted to (a combination of) other systems such as cold atoms \cite{wigley2015fast}.

\begin{figure}[t]
\centerline{\includegraphics[width=0.48 \textwidth]{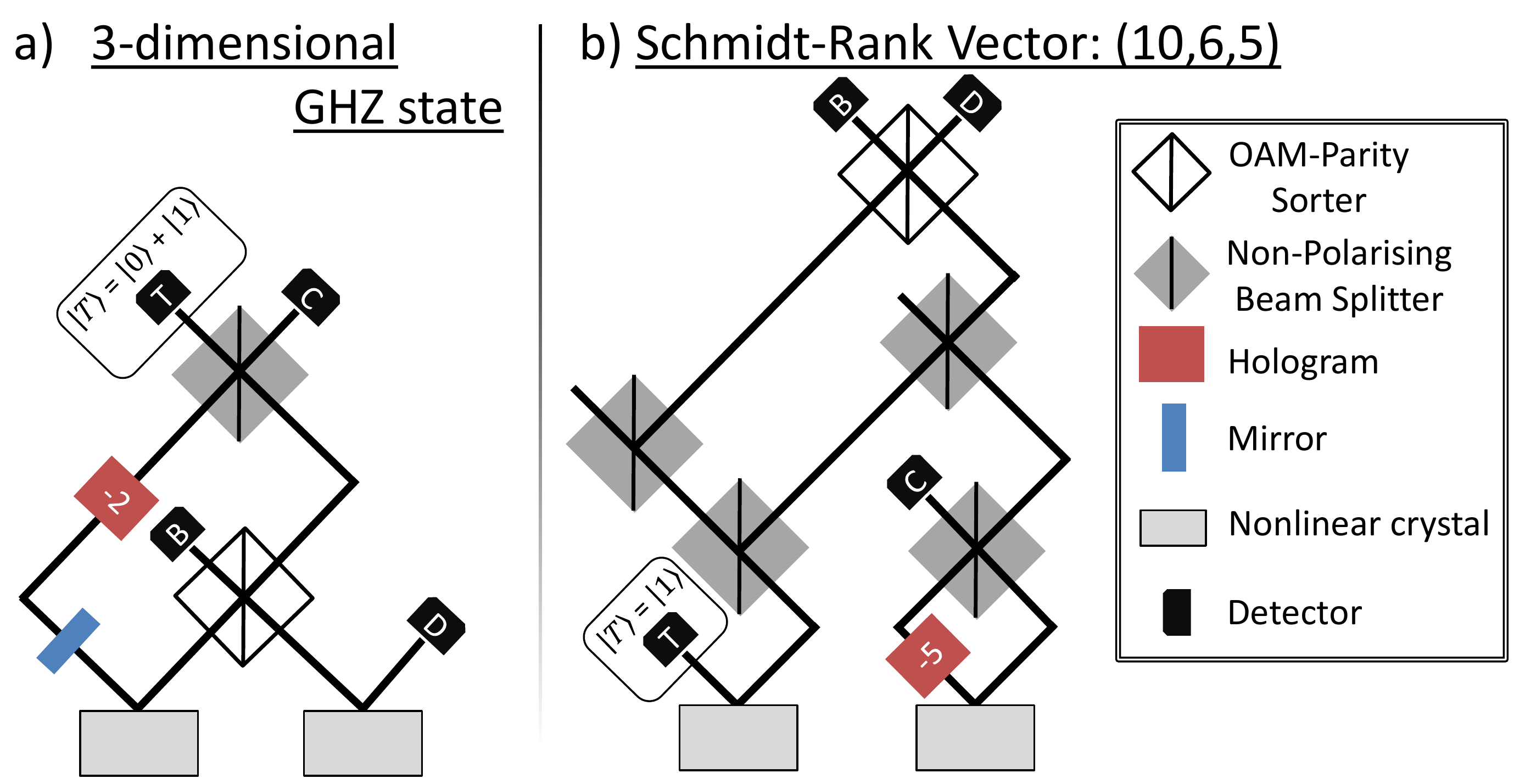}}
\caption{Experimental implementations of high-dimensional multipartite entangled quantum states. \textbf{a}) The experimental implementation for a 3-dimensional 3-partite GHZ-State. If Detector T (Trigger) observes a photon in the state $\ket{T}=\left( \ket{0} + \ket{1} \right)$, then the rest of the quantum state is in a GHZ-state, which looks like $\ket{\psi}=\ket{0,0,0}+\ket{1,1,1}+\ket{2,2,2}$ (up to  local transformations). The parity sorter, as described in \cite{leach2002measuring}, can sort even and odd OAM modes. The 3-dimensional GHZ-state has a Schmidt-Rank-Vector of (3,3,3) (all components are symmetrically entangled with the rest of the state). \textbf{b}) A more complex experiment is required for higher-order Schmidt-Rank-Vectors. The (10,6,5)-state is one example of asymmetrically entangled quantum states. The experiments are just two examples of 51 implementations found for creating a variety of different entangled states.}
\label{fig:multipartiteStates}
\end{figure}
\textit{Example 1: High-dimensional multipartite entanglement} - The Greenberger-Horne-Zeilinger (GHZ) state is the most prominent example for non-classical correlations between more than two involved parties, and has led to new understanding of the fundamental properties of quantum physics \cite{greenberger1989going}. It has been shown recently that its generalization to higher dimensions not only has curious properties \cite{lawrence2014rotational}, but that it is a limiting case of a much richer class of non-classical correlations \cite{huber2013structure, huber2013entropy, cadney2014inequalities}. Those new structures of multipartite high-dimensional entanglement are characterized by the Schmidt-Rank Vector and give rise to new phenomena that only exist if both the number of particles and the number of dimensions are larger than two. An example of a state with Schmidt-Rank Vector (4,2,2) is the asymmetrically entangled state $\ket{\psi_{4,2,2}}=\frac{1}{2}\left(\ket{0,0,0}+\ket{1,0,1}+\ket{2,1,0}+\ket{3,1,1}\right)$. There, the first particle is 4-dimensionally entangled with the other two parties, whereas particle 2 and 3 are both only two-dimensionally entangled with the rest. This Master-Slave-Slave configuration is one of the yet unexplored features that only exist in genuine high-dimensional multipartite entanglement, and will be interesting to study in more detail in future. In order to make future experimental investigations possible, we aim to find high-dimensional multipartite entangled states in photonic systems.

Here, the initial state is created by a double spontaneous parametric down-conversion process (SPDC). SPDC is a widespread source for experimental generation of photon pairs. Multiple SPDC processes can produce multipartite entanglement, as it is well-known for the case of two-dimensional polarization entanglement \cite{bouwmeester1999observation, yao2012observation}. However, instead of polarization, we use the orbital angular momentum (OAM) of photons \cite{allen1992orbital, dada2011experimental, romero2012increasing, krenn2014generation}, which is a discrete high-dimensional degree of freedom based on the spatial structure of the photonic wave function. 

The experiments are generated using a set of basic elements consisting of beam splitters, mirrors, dove prism, holograms and OAM-parity sorters \cite{leach2002measuring, Supplementary}. The holograms and the dove prisms have discrete parameters corresponding to the OAM and phase added to the beam, respectively. These elements are randomly placed in one out of six different paths (four of the paths are inputs of the two photon pairs and two are empty to increase variability). One arm is used to trigger the tri-partite state in the other three arms, which leads to roughly $10^{15}$ possible configurations. At the end, a post-selection procedure consisting of the coincidence detection of four photons in the first four arms yields the final state.

We calculate the Schmidt-Rank-Vector of the final state and select non-trivial ones (i.e. where there are no separable parties). Furthermore, for higher usefulness in experiments, we demand that the final state is maximally entangled in its orbital-angular-momentum. If the criteria hold, the experiment is reported.

\melvin runs for roughly 150 hours (on an Intel Core i7 notebook with 2,4 GHz and 24 GB RAM), and finds 51 experiments for states that are entangled in genuinely different ways. Among them, we find the first experimentally realizable scheme of a high-dimensional GHZ-state \cite{lawrence2014rotational}, a generalization of the well-studied two-dimensional GHZ-state (fig. \ref{fig:multipartiteStates}A). Furthermore, we find many experiments for different asymmetrically entangled states (such as the $\ket{\psi_{4,2,2}}$ explained above). In addition, several experiments only differ by continuously tunable components (e.g. different holograms or triggers), making it possible to explore continuous transitions between states of different classes of entanglement.

The resulting experiments contain interesting novel experimental techniques previously unknown to the authors. For example, in 50 out of 51 experiments, one of the four paths that comes directly from the crystals has not been mixed with any other arm (arm D in fig. \ref{fig:multipartiteStates}A, and arm T in \ref{fig:multipartiteStates}B). The reason is that for double SPDC events it is possible that the two photon pairs come from the same crystal. Leaving one path unmixed leads to erasure of such double-pair emission events in four-fold coincidence detection. Interestingly, this immediately introduces asymmetry in the final experimental configuration. A different novelty is introduced when more than 6-dimensional entanglement is created beginning from two three-dimensional entangled pairs. This is only possible when the OAM spectra in two crystals are shifted with a hologram and combined in a nontrivial way (a preliminary stage of the technique can be seen in fig. \ref{fig:multipartiteStates}B, where the spectrum in arm C is shifted in order to reach a 10-dimensional output). In other experiments, the normalization of the state has to be adjusted in order to get a maximally entangled output. As neutral-density filters were not part of the toolbox, \melvin instead used beam-splitters as a 50\%-filter (for example, fig. \ref{fig:multipartiteStates}B).

Now we briefly explain the 3-dimensional GHZ-state experiment (fig. \ref{fig:multipartiteStates}A, details in \cite{Supplementary}): Two independent SPDC events in two crystals (which produce 3-dimensional entangled pairs) allow for nine different states in the four arms. The parity-sorter effectively removes all combinations with opposite OAM-parity from two crystals (such as $\ket{0,0,-1,+1}$), which reduces the state to five terms. Detection photon A in the trigger state $\ket{T}=\ket{0}+\ket{1}$) leads to a multipartite entangled state where photons C and D reside in a three-dimensional space and photon B lives in a two-dimensional space \cite{malik2015multi}. The dimensionality of photon B is then increased from 2 to 3 in an intricate combination of photons A and C. Photon A is shifted by -2 OAM quanta and combined with photon C at a beam splitter. These photons are then detected in the same mode in one BS output, which effectively erases the "which-crystal" information and entangles the remaining three photons into a 3-dimensional GHZ state.

\begin{figure}[t]
\centerline{\includegraphics[width=0.4 \textwidth]{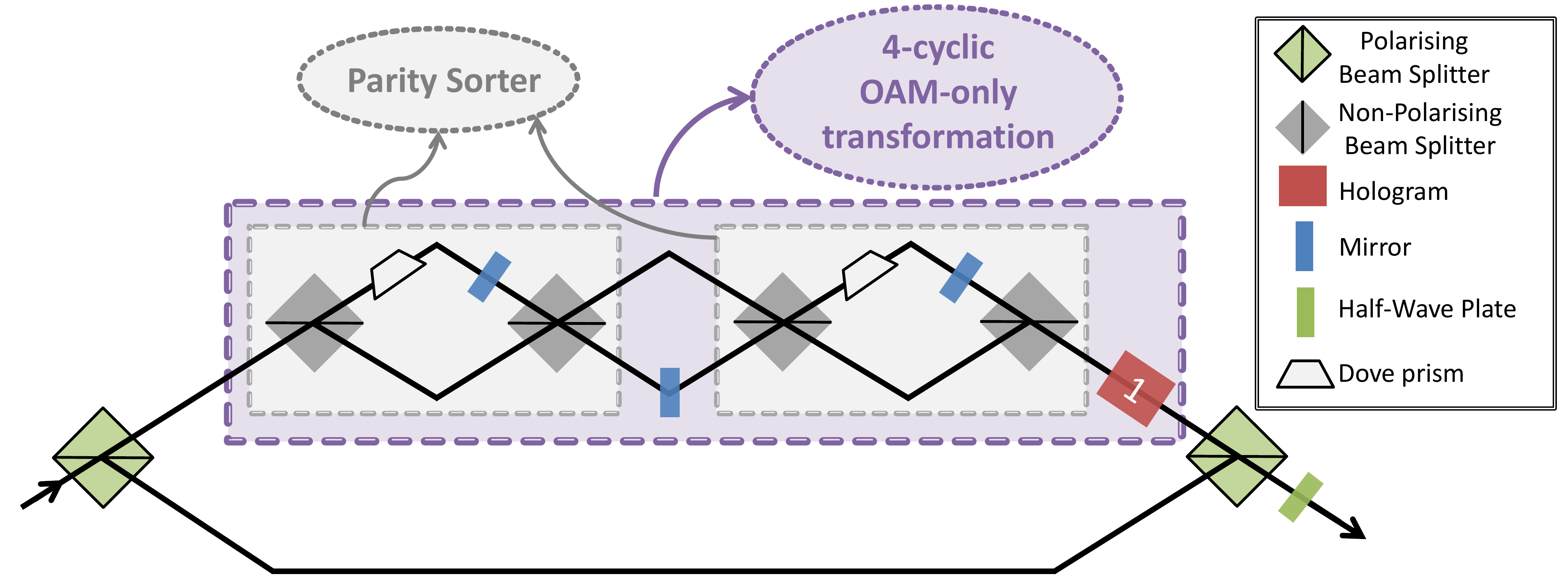}}
\caption{Realization of an 8-cyclic rotation using polarization and OAM ($\ket{-1,H} \to \ket{-1,V} \to \ket{0,V} \to ... \ket{2,H} \to \ket{-1,V}$). In the experiment, a 4-cyclic rotation for pure OAM values is used. Within the 4-cyclic rotation, the parity sorter [29] mentioned in the main text is used twice.}
\label{fig:cyclic}
\end{figure}
\textit{Example 2: High-dimensional cyclic operations and learning} - In the second example, we are interested in high-dimensional cyclic rotations, which are special cases of high-dimensional unitary transformations. A set of states is transformed in such a way that the last element of the set transforms to the first element (for example, $\ket{1} \to \ket{2} \to \ket{3} \to \ket{1}$ is a 3-cycle). Such transformations are required in novel kinds of high-dimensional quantum information protocols \cite{araujo2014computational, tavakoli2015secret} as well as in the creation of high-dimensional Bell-states. Here, our input is a set of high-dimensional states encoded in different degrees of freedom (path, polarization, and OAM). While the creation and verification of high-dimensional entanglement in OAM is well known \cite{romero2012increasing, krenn2014generation}, the knowledge of how to perform arbitrary transformations in this degree-of-freedom is still lacking. Thus, finding such transformations in OAM is very important, as it would enable practical experiments with high-dimensional quantum states and find application in high-dimensional quantum information protocols.

The experiments are generated using a set of basic elements that consists of polarizing and non-polarizing beam splitters, dove prisms, mirrors, holograms and half-wave plates. These elements are placed in one of three different paths (one path is used as an input, and two empty paths are added to increase variability). This leads to roughly $10^{22}$ different possible experimental configurations.

The criterion is based on the largest cycle of the transformation: A number of input states (with different polarization (horizontal and vertical), OAM ($\ell$=-10 to +10) and paths) is calculated. Then we search subsets of modes that are transformed in a closed cycle, as described above, and select the largest closed cycle. \melvin was able to find the first experimentally realizable OAM-only 4-cyclic transformation, OAM and polarization 3-, 6-, and 8-cyclic rotations and up to 14-cyclic rotations using OAM, polarization and path (fig. \ref{fig:cyclic} \& \cite{Supplementary}). 

\begin{figure}[t]
\centerline{\includegraphics[width =0.35 \textwidth]{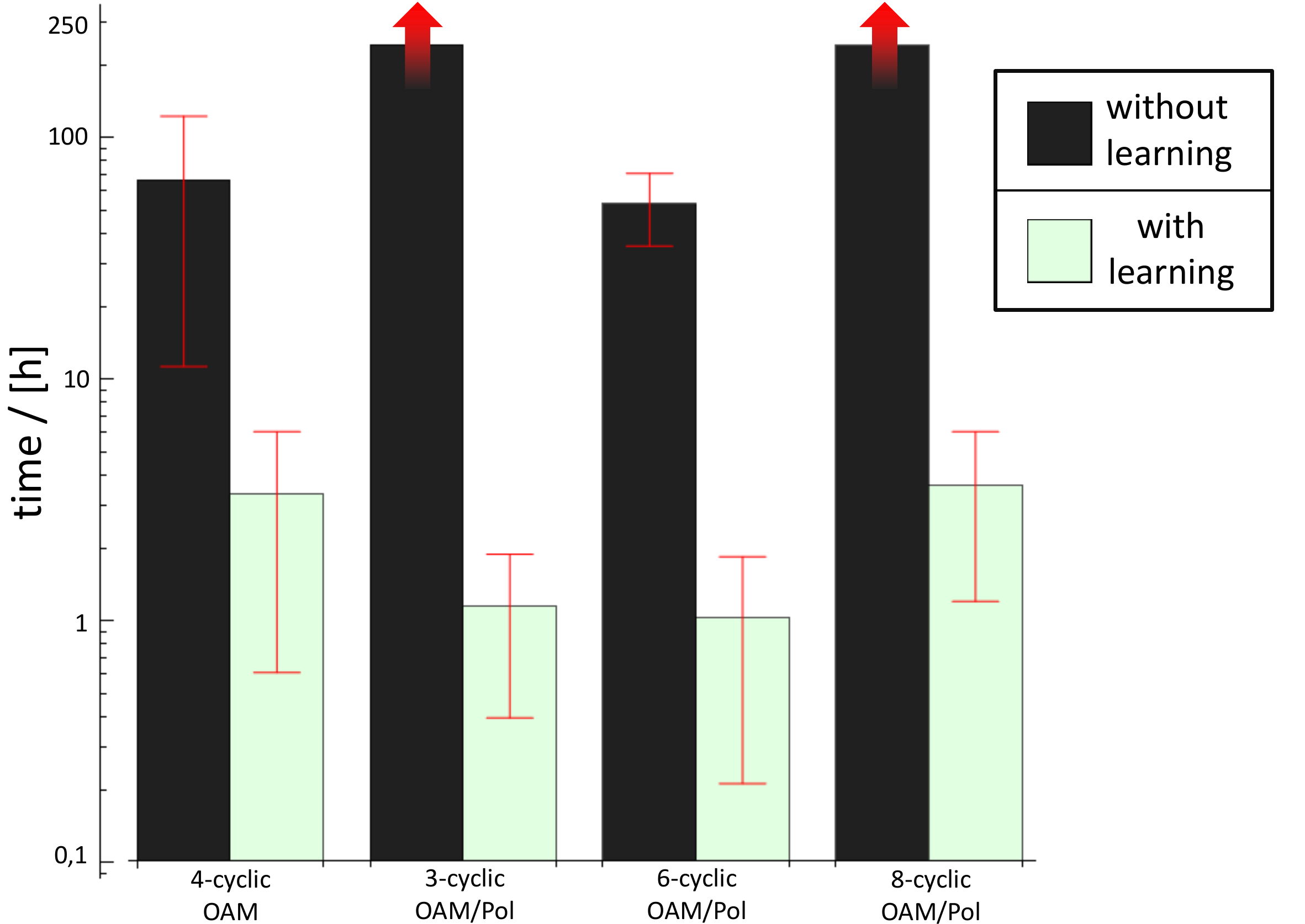}}
\caption{Comparison of performance with and without the ability to learn (log-scale). Green shows the average time required in the case where the algorithm can learn useful transformations (the algorithm was executed 10 times with the same initial conditions). Black shows the time it requires without the ability to learn. The experiments for 3-cyclic and 8-cyclic transformations were not found (within 250 hours) without learning, while experiments for 4-cyclic and 6-cyclic rotation were found three and four times in 250 hours, respectively. The errors stand for one standard deviation, calculated from the times it took to find the solution. Thus, autonomously extending the set of useful transformations improves \melvin's performance, which is crucial for scaling to more complex experiments. }
\label{fig:learning}
\end{figure}
Complex problems can be solved more efficiently by reusing solutions to simpler problems \cite{schmidt2009distilling, briegel2012projective}: Whenever \melvin finds a solution for a simpler system, it memorizes the experimental configuration as new part of its initial toolbox \cite{Supplementary}. The novel elements in the toolbox can be used to construct the next experimental configuration. To compare the effectiveness of learning, we analyze the algorithm with and without the ability to increase its own set of basic elements. We ran the algorithm for 250 hours, and only 3 and 4 instances of 4-cyclic and 6-cyclic rotations were found, respectively. Not a single instance of a 3-cyclic and an 8-cyclic rotation was found within 250 hours. However, using the ability to learn new elements we ran the algorithm 10 times (starting with the initial toolbox, i.e. without keeping the learned elements), and discovered that the 3- and 6-cyclic rotations were found on average within 90 minutes (they were always found within 3 hours), and the 4- and 8- cyclic rotations were found on average within 3,5 hours (in each of the 10 trials, they were found within 8 hours). Thus the ability to learn new elements improves the search by more than one order of magnitude, suggesting a mechanism for experiments with a higher complexity (fig. \ref{fig:learning}).

\textit{Conclusion and Outlook} - We have shown how a computer can find new quantum experiments. The large number of discoveries reveals a way to investigate new families of complex entangled quantum systems in the laboratory. Several of these experiments are being built at the moment in our labs \cite{malik2015multi, schlederer2015cyclic}. In contrast to human designers of experiments, \melvin does not follow intuitive reasoning about the physical system, and therefore leads to the utilization of many unfamiliar and unconventional techniques that are challenging to understand. The algorithm can learn from experience (i.e. previous successful solutions), which leads to a significant speed-up in discoveries of more complex experiments. 

\melvin can be applied to many other questions about the creation and manipulation of quantum systems, such as the search for more general high-dimensional transformations with different degrees-of-freedom and for different physical systems such as ultra-cold atoms \cite{wigley2015fast}, or for efficient generation of other types of important quantum systems such as NOON-states \cite{afek2010high}. In order to improve the efficiency of finding solutions, powerful techniques from artificial intelligence research can be applied, such as evolutionary algorithms \cite{eiben2015evolutionary} (where the experiment and the resulting quantum state play the role of genotype and phenotype, respectively), reinforcement learning techniques \cite{briegel2012projective, littman2015reinforcement, mnih2015human} (by implementing a reward-function depending on the closeness of the quantum state's properties to the desired properties) or entropy-based \cite{wissner2013causal} and big-data methods \cite{varshney2013big} (in order to find more unexpected solutions).

\textit{Acknowledgement} - We thank Daniel Greenberger, Marcus Huber, Hans Briegel, Nora Tischler and Dominik Leitner for fruitful discussions and/or critical reading of our manuscript. This project was supported by the Austrian Academy of Sciences (\"OAW), the European Research Council (SIQS Grant No. 600645 EU-FP7-ICT), the Austrian Science Fund (FWF) with SFB F40 (FOQUS). MM acknowledges support from the European Commission through a Marie Curie fellowship (OAMGHZ).

\bibliography{refs}

\onecolumngrid
\vspace{\columnsep}
\input{mvSupp.tex}

\end{document}

%% file: mvSupp.tex

\begin{center}\LARGE{Supplementary Informations}
\end{center}

\subsection{S1) High-dimensional GHZ-entanglement}
A straight-forward generalization of a 3-partite GHZ-state to 3 dimensions looks like
\begin{eqnarray}
\ket{\psi}=\frac{1}{\sqrt{3}} \left( \ket{0,0,0} + \ket{1,1,1} + \ket{2,2,2} \right).
\label{eq:3dimGHZ1}
\end{eqnarray}
Local transformations do not change the properties of entanglement, thus a 3-dimensional 3-partite GHZ-state can be written as
\begin{eqnarray}
\ket{\psi}=\frac{1}{\sqrt{3}}\left( \ket{a,b,c} + \ket{\bar{a},\bar{b},\bar{c}} + \ket{\bar{\bar{a}},\bar{\bar{b}},\bar{\bar{c}}} \right),
\label{eq:3dimGHZ2}
\end{eqnarray}
where $x \perp \bar{x} \perp \bar{\bar{x}}$ with $x = \{a,b,c\}$.

\subsection{S2) Spontaneous parametric down-conversion (SPDC)}
The input state of example 1 is a double-emission from SPDC, in the form of
\begin{eqnarray}
\ket{\psi_{SPDC}}=N \left(\sum_{\ell=-DC}^{DC} \ket{+\ell_A, -\ell_B} + \ket{+\ell_C, -\ell_C} \right)^2
\label{eq:SPDC1}
\end{eqnarray}
with DC being the highest order of SPDC considered, and A, B, C and D are the path of the photons, and N is a normalization constant. We post-select on fourfold coincidences. Such a state can be written with DC=1 gives
\begin{eqnarray}
\ket{\psi_{SPDC,AB+CD}}=N \left(\ket{0_A,0_B} + \ket{-1_A,1_B} + \ket{1_A,-1_B} + \ket{0_C,0_D} + \ket{-1_C,1_D} + \ket{1_C,-1_D} \right)^2
\label{eq:SPDC1}
\end{eqnarray}
In the examples, we consider DC=1 up to DC=3 (see chapter S4) to generate the state, and we make sure that up to DC=25 higher-order terms don't modify the post-selected output.
\subsection{S3) Basis elements from the toolbox}
Here we list the symbolic transformations of all elements used in the toolbox. Here, $a$ or $b$ stand for the path of the photon, $\ell$ and $P$ stand for the photon's OAM and polarization.

\subsubsection{Reflection}

\begin{eqnarray}
\textnormal{Reflection}[\psi,\TN{a}] = \psi \Leftarrow \left\{
  \begin{array}{lr}
    \TN{a}[\ell,H]\rightarrow -i\cdot\TN{a}[-\ell,H]\\
    \TN{a}[\ell,V]\rightarrow i\cdot\TN{a}[-\ell,V]
  \end{array}
\right.
\label{eq:symbRef}
\end{eqnarray}

\subsubsection{Non-polarizing symmetric 50/50 beam splitter}
\begin{eqnarray}
\textnormal{BS}[\psi,\TN{a},\TN{b}] = \psi \Leftarrow \left\{
  \begin{array}{lr}
    \TN{a}[\ell,P]\rightarrow \frac{1}{\sqrt{2}}\left(\TN{b}[\ell] + \TN{Reflection}[\TN{a}[\ell,P]]\right)\\
    \TN{b}[\ell,P]\rightarrow \frac{1}{\sqrt{2}}\left(\TN{a}\TN[\ell] + \TN{Reflection}[\TN{b}[\ell,P]]\right)
  \end{array}
\right.
\label{eq:symbBSFull}
\end{eqnarray}

\subsubsection{Polarizing beam splitter}
\begin{eqnarray}
\textnormal{PBS}[\psi,\TN{a},\TN{b}] = \psi \Leftarrow \left\{
  \begin{array}{lr}
    \TN{a}[\ell,H]\rightarrow \TN{b}[\ell,H]\\
    \TN{a}[\ell,V]\rightarrow i\cdot\TN{a}[-\ell,V]\\    
    \TN{b}[\ell,H]\rightarrow \TN{a}[\ell,H]\\
    \TN{b}[\ell,V]\rightarrow i\cdot\TN{b}[-\ell,V]     
  \end{array}
\right.
\label{eq:symbPBS}
\end{eqnarray}

\subsubsection{Half-Wave Plate}
\begin{eqnarray}
\textnormal{HWP}[\psi,\TN{a}] = \psi \Leftarrow \left\{
  \begin{array}{lr}
    \TN{a}[\ell,H]\rightarrow \TN{a}[\ell,V]\\
    \TN{a}[\ell,V]\rightarrow -\TN{a}[\ell,H]    
  \end{array}
\right.
\label{eq:symbHWP}
\end{eqnarray}

\subsubsection{OAM hologram}
\begin{eqnarray}
\textnormal{OAMHolo}[\psi,\TN{a},\TN{n}] = \psi \Leftarrow \left\{
    \TN{a}[\ell,P]\rightarrow \TN{a}[\ell + n,P]
    \right.    
\label{eq:symbHolo}
\end{eqnarray}

\subsubsection{OAM hologram-superposition}
\begin{eqnarray}
\textnormal{OAMHoloSP}[\psi,\TN{a},\TN{n}] = \psi \Leftarrow \left\{
    \TN{a}[\ell,P]\rightarrow \frac{1}{\sqrt{2}} \left(\TN{a}[\ell,P] + \TN{a}[\ell + n,P] \right)
    \right.    
\label{eq:symbHoloSP}
\end{eqnarray}

\subsubsection{Dove prism}
\begin{eqnarray}
\textnormal{DP}[\psi,\TN{a},\TN{n}] = \psi \Leftarrow \left\{
    \TN{a}[\ell,P]\rightarrow e^{i\frac{\pi}{n}\ell} \TN{Reflection}[\TN{a}[\ell,P]] 
    \right.    
\label{eq:symbDP}
\end{eqnarray}

\subsubsection{OAM Parity Sorter}
\begin{eqnarray}
\TN{LI}[\psi,\TN{a},\TN{b}]= \TN{BS}[\TN{Reflection}[\TN{Reflection}[\TN{DP}[\TN{Reflection}[\TN{BS}[\psi,\TN{a},\TN{b}],\TN{a}],\TN{a},1],\TN{b}],\TN{b}],\TN{a},\TN{b}]
\label{eq:symbLI}
\end{eqnarray}

\subsection{S4) Experimental Implementations for High-dimensional 3-partite Entanglement}
High-dimensional multipartite entanglement can be characterized by the Schmidt-Rank Vector, introduced in \cite{huber2013structure}. In example 1, we search for such states, and find many examples, as shown in fig.(\ref{fig:SRVtable}) below.
\begin{figure}[t]
\centerline{\includegraphics[width =0.38 \textwidth]{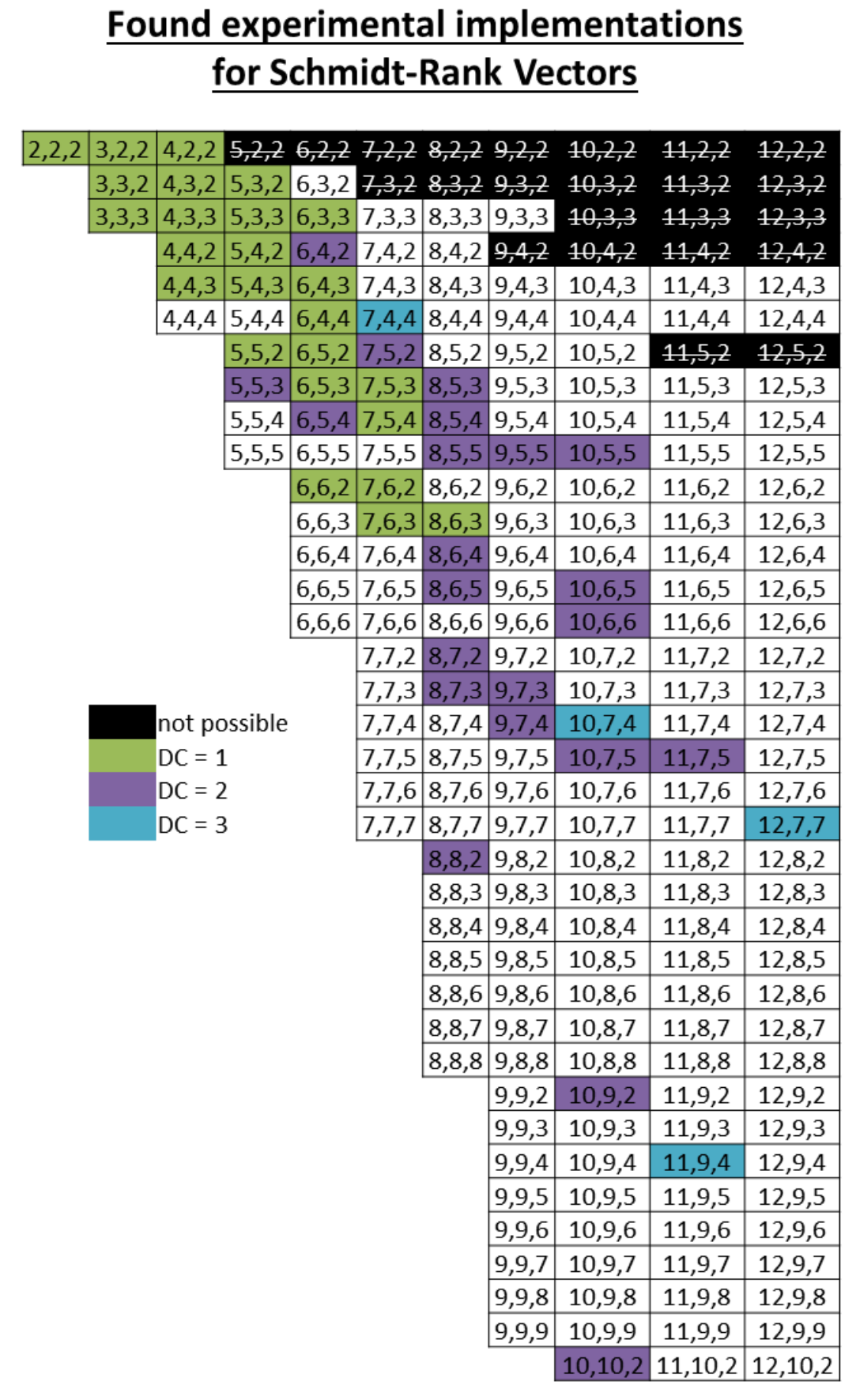}}
\caption{Found experimental implementations for high-dimensional three-partite entangled states. The list shows experimental setups with states of different Schmidt-Rank Vectors. Black cells are not possible \cite{cadney2014inequalities}. Cells with green, violet and blue filling indicate states that can be generated with 1st, 2nd or 3rd order in the OAM spectrum (see chapter S2). For white cells, no experimental realization has been found yet.}
\label{fig:SRVtable}
\end{figure}
All Schmidt-Rank-vectors (SRV), their experiments, the required trigger and the resulting quantum state are listed here:

\texttt{\scriptsize{
\begin{center}
  \begin{tabular}{|c|l|l|l|}
    \hline
    \textbf{SRV} & \textbf{Experimental Setup} & \textbf{Trigger in A} & \textbf{resulting quantum state}\\ \hline
    \textbf{DC=1} &   &   &  \\ \hline    
    (2,2,2) & "OAMHolo[$\psi$,c,-1]", "LI[XXX,a,c]" & $\ket{1}+\ket{2}$ & \pbox{20cm}{-FF2[-1]*FF3[-1]*FF4[0] - FF2[0]*FF3[0]*FF4[1]} \\ \hline
    (3,2,2) & {"BS[$\psi$,b,c]", "LI[XXX,c,f]"} & $\ket{1}$ & \pbox{20cm}{-FF2[0] FF3[-1] FF4[0] - FF2[-1] FF3[-1] FF4[1] +\\ + FF2[1] FF3[1] FF4[1]} \\ \hline    
    (3,3,2) & {"LI[$\psi$,b,c]"} & $\ket{-1} + \ket{0}$ & \pbox{20cm}{-FF2[1]*FF3[1]*FF4[-1] - FF2[0]*FF3[0]*FF4[0] - \\ - FF2[1]*FF3[-1]*FF4[1]} \\ \hline
    (3,3,3) & \pbox{20cm}{"LI[$\psi$,b,c]", "Reflection[XXX,a]",\\"OAMHolo[XXX,a,-2]", "BS[XXX,a,c]"} & $\ket{1} + \ket{0}$ & \pbox{20cm}{-I FF2[-1] FF3[-3] FF4[-1] - I FF2[0] FF3[-2] FF4[0] + \\ + I FF2[1] FF3[1] FF4[1]} \\ \hline
    (4,2,2) & \pbox{20cm}{"OAMHolo[$\psi$,b,-1]", "BS[XXX,a,d]",\\"BS[XXX,b,d]", "LI[XXX,d,f]"} & $\ket{0}$ & \pbox{20cm}{FF2[-1] FF3[-1] FF4[-1] + FF2[2] FF3[0] FF4[-1] - \\ - FF2[1] FF3[-1] FF4[1] + FF2[0] FF3[0] FF4[1]} \\ \hline
    (4,3,2) & \pbox{20cm}{"OAMHoloSP2[$\psi$,c,-3]", "LI[XXX,a,c]"} & $\ket{1} + \ket{2}$ &  \pbox{20cm}{-FF2[0]*FF3[0]*FF4[-1] - FF2[-1]*FF3[1]*FF4[-1] - \\ - FF2[-1]*FF3[-3]*FF4[0] - FF2[-1]*FF3[-1]*FF4[1]} \\ \hline
    (4,3,3) & \pbox{20cm}{"BS[$\psi$,c,f]", "OAMHolo[XXX,c,1]",\\"OAMHolo[XXX,a,2]", "BS[XXX,c,f]",\\ "BS[XXX,a,d]", "LI[XXX,c,d]"} & \pbox{20cm}{$\ket{1} + \ket{0}$ + \\ $\ket{-1}$} & \pbox{20cm}{FF2[1]*FF3[-1]*FF4[-1] + FF2[1]*FF3[0]*FF4[0] + \\ + FF2[0]*FF3[-2]*FF4[2] - FF2[-1]*FF3[1]*FF4[3]} \\ \hline
    (5,3,2) & \pbox{20cm}{"OAMHolo[$\psi$,b,-4]", "BS[XXX,b,c]", \\ "LI[XXX,d,f]", "BS[XXX,a,c]"} & $\ket{-1}$ & \pbox{20cm}{FF2[3]*FF3[-1]*FF4[-1] + FF2[4]*FF3[0]*FF4[-1] + \\ + FF2[1]*FF3[5]*FF4[-1] - FF2[5]*FF3[-1]*FF4[1] + \\ + FF2[-1]*FF3[5]*FF4[1]} \\ \hline
    (5,3,3) & \pbox{20cm}{"OAMHoloSP2[$\psi$,a,5]", "LI[XXX,a,c]"} & $\ket{0} + \ket{-1}$ & \pbox{20cm}{-FF2[1]*FF3[1]*FF4[-1] - FF2[-1]*FF3[-6]*FF4[0] - \\ - FF2[1]*FF3[-4]*FF4[0] - FF2[0]*FF3[0]*FF4[0] - \\ - FF2[1]*FF3[-1]*FF4[1]} \\ \hline
    (5,4,2) & \pbox{20cm}{"OAMHolo[$\psi$,d,1]", "OAMHolo[XXX,c,-3]", \\ "LI[XXX,a,d]", "BS[XXX,b,c]", \\ "LI[XXX,a,c]"} & $\ket{0} + \ket{1}$  & \pbox{20cm}{FF2[0]*FF3[-1]*FF4[0] + FF2[-2]*FF3[0]*FF4[0] + \\ + FF2[-4]*FF3[2]*FF4[0] + FF2[-3]*FF3[-1]*FF4[1] - \\ - FF2[1]*FF3[3]*FF4[1]} \\ \hline
    (5,4,3) & \pbox{20cm}{"OAMHolo[$\psi$,a,3]", "BS[XXX,a,b]",\\ "LI[XXX,a,d]"} & $\ket{0} + \ket{1}$ & \pbox{20cm}{-(FF2[2]*FF3[1]*FF4[-1]) - FF2[3]*FF3[0]*FF4[0] - \\ - FF2[2]*FF3[-1]*FF4[1] + FF2[-1]*FF3[0]*FF4[2] + \\ + FF2[1]*FF3[0]*FF4[4]} \\ \hline
    (5,5,2) & \pbox{20cm}{"OAMHolo[$\psi$,c,-4]", "BS[XXX,c,d]",\\ "LI[XXX,a,d]"} & $\ket{0} + \ket{1}$ & \pbox{20cm}{-(FF2[-1]*FF3[1]*FF4[-5]) - FF2[-1]*FF3[-1]*FF4[-3] + \\ + FF2[-1]*FF3[5]*FF4[-1] + FF2[0]*FF3[4]*FF4[0] + \\ + FF2[-1]*FF3[3]*FF4[1]} \\ \hline
    (6,3,3) & \pbox{20cm}{"OAMHolo[$\psi$,a,4]", "Reflection[XXX,b]", \\ "BS[XXX,a,d]", "LI[XXX,a,e]", \\ "BS[XXX,a,b]"} & $\ket{-1}$ & \pbox{20cm}{FF2[-3]*FF3[-1]*FF4[-1] + FF2[-3]*FF3[0]*FF4[0] +  \\ + FF2[-3]*FF3[1]*FF4[1] - FF2[-1]*FF3[1]*FF4[3] +  \\ + FF2[0]*FF3[-1]*FF4[4] + FF2[-1]*FF3[-1]*FF4[5]} \\ \hline
    (6,4,3) & \pbox{20cm}{"OAMHolo[$\psi$,a,-4]", "BS[XXX,a,d]", \\ "LI[XXX,a,c]"} & $\ket{0}+\ket{3}$ & \pbox{20cm}{-(FF2[1]*FF3[0]*FF4[-5]) - FF2[0]*FF3[0]*FF4[-4] - \\ - FF2[-1]*FF3[0]*FF4[-3] + FF2[-1]*FF3[-1]*FF4[-1] + \\ + FF2[0]*FF3[-4]*FF4[0] + FF2[-1]*FF3[1]*FF4[1]} \\ \hline
  \end{tabular}
  \begin{tabular}{|c|l|l|l|}
    \hline
    \textbf{SRV} & \textbf{Experimental Setup} & \textbf{Trigger in A} & \textbf{resulting quantum state}\\ \hline
    (6,4,4) & \pbox{20cm}{"OAMHoloSP2[$\psi$,d,5]", "BS[XXX,c,d]", \\ "LI[XXX,b,d]"} & $\ket{0}+\ket{1}$ & \pbox{20cm}{FF2[-1]*FF3[0]*FF4[-5] - FF2[-1]*FF3[6]*FF4[-1] + \\ + FF2[4]*FF3[-1]*FF4[0] + FF2[6]*FF3[1]*FF4[0] - \\ - FF2[0]*FF3[5]*FF4[0] - FF2[-1]*FF3[4]*FF4[1]} \\ \hline
    (6,5,2) & \pbox{20cm}{"OAMHoloSP2[$\psi$,d,6]", "BS[XXX,c,d]", \\ "BS[XXX,a,c]", "LI[XXX,b,f]"} & $\ket{-1}$ & \pbox{20cm}{-FF2[-1]*FF3[-1]*FF4[-7]) - FF2[-1]*FF3[0]*FF4[-6] + \\ + FF2[1]*FF3[-1]*FF4[-5] + FF2[-1]*FF3[-7]*FF4[-1] + \\ + FF2[-1]*FF3[-6]*FF4[0] + FF2[-1]*FF3[-5]*FF4[1]} \\ \hline
    (6,5,3) & \pbox{20cm}{"OAMHolo[$\psi$,c,6]", "BS[XXX,c,d]", \\ "LI[XXX,b,c]"} & $\ket{-1}+\ket{0}$ & \pbox{20cm}{FF2[1]*FF3[-5]*FF4[-1] + FF2[6]*FF3[0]*FF4[0] + \\ + FF2[1]*FF3[-7]*FF4[1] - FF2[1]*FF3[1]*FF4[5] - \\ - FF2[0]*FF3[0]*FF4[6] - FF2[1]*FF3[-1]*FF4[7]} \\ \hline
    (6,6,2) & \pbox{20cm}{"LI[$\psi$,a,d]", "OAMHolo[XXX,c,3]", \\ "BS[XXX,c,d]"} & $\ket{-1}+\ket{0}$ & \pbox{20cm}{FF2[1]*FF3[-2]*FF4[-1] + FF2[0]*FF3[-3]*FF4[0] + \\ + FF2[1]*FF3[-4]*FF4[1] - FF2[1]*FF3[1]*FF4[2] - \\ - FF2[0]*FF3[0]*FF4[3] - FF2[1]*FF3[-1]*FF4[4]} \\ \hline
    (7,5,3) & \pbox{20cm}{"OAMHoloSP2[$\psi$,c,4]", "OAMHoloSP2[XXX,a,7]", \\ "BS[XXX,c,d]", "LI[XXX,a,c]"} & $\ket{1} + \ket{4}$ & \pbox{20cm}{FF2[-1]*FF3[-3]*FF4[-1] + FF2[-1]*FF3[-8]*FF4[0] + \\ + FF2[1]*FF3[-6]*FF4[0] + FF2[0]*FF3[0]*FF4[0] + \\ + FF2[-1]*FF3[-5]*FF4[1] - FF2[-1]*FF3[1]*FF4[3] - \\ - FF2[-1]*FF3[-1]*FF4[5]} \\ \hline
    (7,5,4) & \pbox{20cm}{"OAMHolo[$\psi$,d,6]", "OAMHolo[XXX,b,3]", \\ "BS[XXX,a,b]", "BS[XXX,c,d]", \\ "LI[XXX,a,d]"} & $\ket{1}+\ket{0}$ & \pbox{20cm}{-FF2[-4]*FF3[1]*FF4[-7] - FF2[-4]*FF3[-1]*FF4[-5] - \\ - FF2[-1]*FF3[6]*FF4[-4] - FF2[1]*FF3[6]*FF4[-2] + \\ + FF2[-4]*FF3[7]*FF4[-1] + FF2[-3]*FF3[6]*FF4[0] + \\ + FF2[-4]*FF3[5]*FF4[1]} \\ \hline
    (7,6,2) & \pbox{20cm}{"OAMHoloSP2[$\psi$,c,-8]", "BS[XXX,c,d]", \\ "OAMHoloSP2[XXX,c,-5]", "LI[XXX,a,c]"} & $\ket{1} + \ket{-4}$ & \pbox{20cm}{-FF2[0]*FF3[0]*FF4[-9] - FF2[1]*FF3[1]*FF4[-9] - \\ - FF2[1]*FF3[-5]*FF4[-8] - FF2[1]*FF3[-1]*FF4[-7] + \\ + FF2[1]*FF3[9]*FF4[-1] + FF2[1]*FF3[3]*FF4[0] + \\ + FF2[1]*FF3[7]*FF4[1]} \\ \hline
    (7,6,3) & \pbox{20cm}{"OAMHolo[$\psi$,c,5]", "OAMHolo[XXX,a,-2]", \\ "BS[XXX,c,d]", "BS[XXX,a,d]"} & $\ket{1}$ & \pbox{20cm}{-FF2[-1]*FF3[-1]*FF4[-6] - FF2[-1]*FF3[0]*FF4[-5] - \\ - FF2[-1]*FF3[1]*FF4[-4] - FF2[1]*FF3[-6]*FF4[-3] - \\ - FF2[0]*FF3[-6]*FF4[-2] + FF2[-1]*FF3[-5]*FF4[0] + \\ + FF2[-1]*FF3[-4]*FF4[1]} \\ \hline
    (8,6,3) & \pbox{20cm}{"OAMHoloSP2[$\psi$,a,6]", "BS[XXX,a,b]", \\ "OAMHoloSP2[XXX,a,-3]", "LI[XXX,a,d]"} & $\ket{0}+\ket{-5}$ & \pbox{20cm}{FF2[-1]*FF3[1]*FF4[-1] - FF2[6]*FF3[0]*FF4[0] + \\ + FF2[-1]*FF3[-1]*FF4[1] - FF2[5]*FF3[0]*FF4[2] - \\ - FF2[7]*FF3[0]*FF4[4] + FF2[0]*FF3[0]*FF4[6] + \\ + FF2[-1]*FF3[0]*FF4[8] + FF2[1]*FF3[0]*FF4[10]} \\ \hline
    \textbf{DC=2} &   &   &  \\ \hline    
    (5,5,3) & \pbox{20cm}{"Reflection[$\psi$,a]", "LI[XXX,b,c]",  \\ "OAMHolo[XXX,a,2]", "BS[XXX,a,c]"} & $\ket{-1} + \ket{2}$ & \pbox{20cm}{(-I)*FF2[-2]*FF3[0]*FF4[-2] + I*FF2[-1]*FF3[-1]*FF4[-1] - \\ - I*FF2[0]*FF3[0]*FF4[0] - I*FF2[1]*FF3[3]*FF4[1] - \\ - I*FF2[2]*FF3[0]*FF4[2]} \\ \hline
    (6,4,2) & \pbox{20cm}{"OAMHolo[$\psi$,a,3]", "BS[XXX,a,d]",  \\ "OAMHolo[XXX,a,4]", "BS[XXX,a,c]",  \\ "LI[XXX,c,f]"} & $\ket{1}$ & \pbox{20cm}{-FF2[-2]*FF3[1]*FF4[-1] + FF2[0]*FF3[1]*FF4[1] - \\ - FF2[1]*FF3[3]*FF4[2] - FF2[0]*FF3[3]*FF4[3] - \\ - FF2[-1]*FF3[3]*FF4[4] - \\ - FF2[-2]*FF3[3]*FF4[5]} \\ \hline
    (6,5,4) & \pbox{20cm}{"OAMHolo[$\psi$,b,4]", "OAMHolo[XXX,d,3]",  \\ "LI[XXX,b,d]", "OAMHolo[XXX,a,6]",  \\ "BS[XXX,a,b]"} & $\ket{-4}+\ket{5}$ & \pbox{20cm}{FF2[2]*FF3[1]*FF4[-6] - FF2[6]*FF3[-1]*FF4[-4] - \\ - FF2[8]*FF3[-1]*FF4[-2] - FF2[5]*FF3[2]*FF4[1] - \\ - FF2[5]*FF3[0]*FF4[3] - FF2[5]*FF3[-2]*FF4[5]} \\ \hline
    (7,5,2) & \pbox{20cm}{"OAMHolo[$\psi$,d,5]", "BS[XXX,a,d]",  \\ "BS[XXX,a,c]", "LI[XXX,b,f]"} & $\ket{1}$ & \pbox{20cm}{-FF2[-1]*FF3[2]*FF4[-7] - FF2[-1]*FF3[1]*FF4[-6] - \\ - FF2[-1]*FF3[0]*FF4[-5] + FF2[1]*FF3[1]*FF4[-4] - \\ - FF2[-1]*FF3[-2]*FF4[-3] - FF2[1]*FF3[4]*FF4[-1] - \\ - FF2[-1]*FF3[4]*FF4[1]} \\ \hline
    (8,5,3) & \pbox{20cm}{"OAMHolo[$\psi$,b,-5]", "BS[XXX,b,d]",  \\ "OAMHolo[XXX,a,-1]", "BS[XXX,a,d]",  \\ "LI[XXX,d,e]"} & $\ket{1}$ & \pbox{20cm}{FF2[3]*FF3[1]*FF4[-3] - FF2[5]*FF3[-1]*FF4[1] + \\ + FF2[7]*FF3[1]*FF4[1] + FF2[2]*FF3[-2]*FF4[5] + \\ + FF2[1]*FF3[-1]*FF4[5] + FF2[0]*FF3[0]*FF4[5] + \\ + FF2[-1]*FF3[1]*FF4[5] + FF2[-2]*FF3[2]*FF4[5]} \\ \hline    
    (8,5,4) & \pbox{20cm}{"OAMHoloSP2[$\psi$,d,5]", "LI[XXX,a,d]"} & $\ket{-2}+\ket{1}$ & \pbox{20cm}{-FF2[-2]*FF3[-2]*FF4[-2] - FF2[-1]*FF3[1]*FF4[-1] - \\ - FF2[0]*FF3[-2]*FF4[0] - FF2[-1]*FF3[-1]*FF4[1] - \\ - FF2[2]*FF3[-2]*FF4[2] - FF2[-1]*FF3[2]*FF4[3] - \\ - FF2[-1]*FF3[0]*FF4[5] - FF2[-1]*FF3[-2]*FF4[7]} \\ \hline
    (8,5,5) & \pbox{20cm}{"OAMHolo[$\psi$,a,-6]", "OAMHolo[XXX,a,2]",  \\ "BS[XXX,a,c]"} & $\ket{2}$ & \pbox{20cm}{FF2[2]*FF3[-6]*FF4[-2] + FF2[1]*FF3[-5]*FF4[-2] + \\ + FF2[0]*FF3[-4]*FF4[-2] + FF2[-1]*FF3[-3]*FF4[-2] - \\ - FF2[-2]*FF3[-1]*FF4[-1] - FF2[-2]*FF3[0]*FF4[0] - \\ - FF2[-2]*FF3[1]*FF4[1] - FF2[-2]*FF3[2]*FF4[2]} \\ \hline
    (8,6,4) & \pbox{20cm}{"OAMHolo[$\psi$,c,-5]", "BS[XXX,c,d]",  \\ "LI[XXX,a,c]"} & $\ket{-2}+\ket{1}$ & \pbox{20cm}{-FF2[-2]*FF3[-2]*FF4[-7] - FF2[0]*FF3[0]*FF4[-7] - \\ - FF2[2]*FF3[2]*FF4[-7] - FF2[-1]*FF3[1]*FF4[-6] - \\ - FF2[-1]*FF3[-1]*FF4[-4] + FF2[-1]*FF3[7]*FF4[-2] + \\ + FF2[-1]*FF3[5]*FF4[0] + FF2[-1]*FF3[3]*FF4[2]} \\ \hline
    (8,6,5) & \pbox{20cm}{"LI[$\psi$,d,e]", "OAMHolo[XXX,b,7]",  \\ "BS[XXX,b,f]", "LI[XXX,d,f]",  \\ "BS[XXX,b,e]", "OAMHolo[XXX,a,2]",  \\ "BS[XXX,a,d]"} & $\ket{-1}$ & \pbox{20cm}{FF2[8]*FF3[-1]*FF4[-1] - FF2[9]*FF3[1]*FF4[0] - \\ - FF2[7]*FF3[1]*FF4[2] - FF2[6]*FF3[1]*FF4[3] - \\ - FF2[5]*FF3[1]*FF4[4] + FF2[-2]*FF3[-2]*FF4[8] + \\ + FF2[0]*FF3[0]*FF4[8] + FF2[2]*FF3[2]*FF4[8]} \\ \hline
    (8,7,2) & \pbox{20cm}{"Reflection[$\psi$,a]", "OAMHoloSP2[XXX,c,9]",  \\ "LI[XXX,d,e]", "LI[XXX,a,c]", \\ "OAMHoloSP2[XXX,b,-3]", "LI[XXX,b,e]"} & $\ket{-8}+\ket{1}$ & \pbox{20cm}{(-I)*FF2[1]*FF3[1]*FF4[-1] - I*FF2[-1]*FF3[-2]*FF4[1] - \\ - I*FF2[1]*FF3[-1]*FF4[1] - I*FF2[-3]*FF3[0]*FF4[1] - \\ - I*FF2[-5]*FF3[2]*FF4[1] + FF2[2]*FF3[7]*FF4[1] + \\ + FF2[0]*FF3[9]*FF4[1] + FF2[-2]*FF3[11]*FF4[1]} \\ \hline
    (8,7,3) & \pbox{20cm}{"OAMHoloSP2[$\psi$,a,6]", "BS[XXX,a,b]",  \\ "LI[XXX,a,c]"} & $\ket{-5}+\ket{2}$ & \pbox{20cm}{FF2[-1]*FF3[1]*FF4[-1] + FF2[-1]*FF3[-1]*FF4[1] - \\ - FF2[4]*FF3[-2]*FF4[2] - FF2[6]*FF3[0]*FF4[2] - \\ - FF2[8]*FF3[2]*FF4[2] + FF2[-2]*FF3[4]*FF4[2] + \\ + FF2[0]*FF3[6]*FF4[2] + FF2[2]*FF3[8]*FF4[2]} \\ \hline
    (8,8,2) & \pbox{20cm}{"Reflection[$\psi$,d]", "OAMHolo[XXX,a,4]",  \\ "OAMHolo[XXX,c,6]", "LI[XXX,b,c]", \\ "BS[XXX,b,d]"} & $\ket{2}+\ket{3}$ & \pbox{20cm}{(-I)*FF2[2]*FF3[-2]*FF4[-8] - I*FF2[0]*FF3[-2]*FF4[-6] - \\ - I*FF2[-2]*FF3[-2]*FF4[-4] + I*FF2[8]*FF3[-2]*FF4[-2] + \\ + I*FF2[-1]*FF3[7]*FF4[-1] + I*FF2[6]*FF3[-2]*FF4[0] - \\ - I*FF2[1]*FF3[7]*FF4[1] + I*FF2[4]*FF3[-2]*FF4[2]} \\ \hline
    (9,5,5) & \pbox{20cm}{"OAMHoloSP2[$\psi$,c,4]", "BS[XXX,c,d]", \\  "OAMHoloSP2[XXX,a,5]", "LI[XXX,a,c]"} & $\ket{4}+\ket{7}$ & \pbox{20cm}{FF2[-2]*FF3[-3]*FF4[-1] + FF2[-1]*FF3[-6]*FF4[0] + \\ + FF2[1]*FF3[-4]*FF4[0] + FF2[-2]*FF3[-2]*FF4[0] + \\ + FF2[0]*FF3[0]*FF4[0] + FF2[2]*FF3[2]*FF4[0] + \\ + FF2[-2]*FF3[-5]*FF4[1] - FF2[-2]*FF3[1]*FF4[3] - \\ - FF2[-2]*FF3[-1]*FF4[5]} \\ \hline
    (9,7,3) & \pbox{20cm}{"OAMHolo[$\psi$,a,5]", "BS[XXX,a,f]", \\  "OAMHolo[XXX,b,1]", "BS[XXX,b,f]", \\  "BS[XXX,a,c]", "LI[XXX,c,f]"} & $\ket{0}+\ket{3}$ & \pbox{20cm}{-FF2[-3]*FF3[-1]*FF4[-1] + FF2[1]*FF3[-7]*FF4[0] + \\ + FF2[-1]*FF3[-5]*FF4[0] + FF2[-3]*FF3[-3]*FF4[0] - \\ - I*FF2[4]*FF3[-2]*FF4[0] - I*FF2[6]*FF3[0]*FF4[0] + \\ + I*FF2[-2]*FF3[4]*FF4[0] + I*FF2[0]*FF3[6]*FF4[0] - \\ - FF2[-3]*FF3[1]*FF4[1]} \\ \hline
    (9,7,4) & \pbox{20cm}{"OAMHolo[$\psi$,d,3]", "Reflection[XXX,d]", \\  "OAMHolo[XXX,b,-6]", "BS[XXX,a,b]",  \\ "LI[XXX,a,d]"} & $\ket{-5}+\ket{4}$ & \pbox{20cm}{(-I)*FF2[-1]*FF3[-2]*FF4[-5] - I*FF2[-1]*FF3[0]*FF4[-3] + \\ + I*FF2[4]*FF3[-1]*FF4[-2] - I*FF2[-1]*FF3[2]*FF4[-1] + \\ + I*FF2[6]*FF3[-1]*FF4[0] + I*FF2[8]*FF3[-1]*FF4[2] - \\ - I*FF2[-2]*FF3[-1]*FF4[4] - I*FF2[0]*FF3[-1]*FF4[6] - \\ - I*FF2[2]*FF3[-1]*FF4[8]} \\ \hline
  \end{tabular}
  \begin{tabular}{|c|l|l|l|}
    \hline
    \textbf{SRV} & \textbf{Experimental Setup} & \textbf{Trigger in A} & \textbf{resulting quantum state}\\ \hline
    (10,5,5) & \pbox{20cm}{"OAMHolo[$\psi$,b,6]", "BS[XXX,b,c]",  \\ "LI[XXX,c,e]", "BS[XXX,a,c]"} & $\ket{1}$ & \pbox{20cm}{FF2[2]*FF3[-7]*FF4[-2] + FF2[1]*FF3[-7]*FF4[-1] - \\ - FF2[-7]*FF3[1]*FF4[-1] + FF2[0]*FF3[-7]*FF4[0] + \\ + FF2[-1]*FF3[-7]*FF4[1] + FF2[-8]*FF3[-2]*FF4[1] + \\ + FF2[-6]*FF3[0]*FF4[1] + FF2[-5]*FF3[1]*FF4[1] + \\ + FF2[-4]*FF3[2]*FF4[1] + FF2[-2]*FF3[-7]*FF4[2]} \\ \hline
    (10,6,5) & \pbox{20cm}{"OAMHolo[$\psi$,c,-5]", "BS[XXX,c,d]",  \\ "BS[XXX,b,e]", "BS[XXX,b,f]",  \\ "BS[XXX,d,e]", "LI[XXX,b,d]"} & $\ket{1}$ & \pbox{20cm}{FF2[-4]*FF3[-1]*FF4[-1] + FF2[-6]*FF3[1]*FF4[-1] - \\ - FF2[2]*FF3[3]*FF4[-1] - FF2[-1]*FF3[4]*FF4[-1] - \\ - FF2[0]*FF3[5]*FF4[-1] - FF2[-2]*FF3[7]*FF4[-1] - \\ - FF2[-1]*FF3[6]*FF4[1] + FF2[-1]*FF3[-2]*FF4[3] + \\ + FF2[-1]*FF3[0]*FF4[5] + FF2[-1]*FF3[2]*FF4[7]} \\ \hline
    (10,6,6) & \pbox{20cm}{"OAMHolo[$\psi$,d,-5]", "BS[XXX,c,d]",  \\ "LI[XXX,b,d]"} & $\ket{-2}+\ket{1}$ & \pbox{20cm}{-FF2[-2]*FF3[-7]*FF4[-2] - FF2[0]*FF3[-5]*FF4[-2] - \\ - FF2[2]*FF3[-3]*FF4[-2] + FF2[-6]*FF3[-1]*FF4[-2] + \\ + FF2[-4]*FF3[1]*FF4[-2] - FF2[-1]*FF3[-4]*FF4[-1] - \\ - FF2[-1]*FF3[-6]*FF4[1] + FF2[-1]*FF3[2]*FF4[3] + \\ + FF2[-1]*FF3[0]*FF4[5] + FF2[-1]*FF3[-2]*FF4[7]} \\ \hline
    (10,7,5) & \pbox{20cm}{"OAMHoloSP2[$\psi$,d,6]", "BS[XXX,c,d]",  \\ "LI[XXX,b,d]"} & $\ket{-2}+\ket{1}$ & \pbox{20cm}{FF2[-1]*FF3[1]*FF4[-7] + FF2[-1]*FF3[-1]*FF4[-5] + \\ + FF2[4]*FF3[-2]*FF4[-2] + FF2[6]*FF3[0]*FF4[-2] + \\ + FF2[8]*FF3[2]*FF4[-2] - FF2[-2]*FF3[4]*FF4[-2] - \\ - FF2[0]*FF3[6]*FF4[-2] - FF2[2]*FF3[8]*FF4[-2] - \\ - FF2[-1]*FF3[7]*FF4[-1] - FF2[-1]*FF3[5]*FF4[1]} \\ \hline
    (10,9,2) & \pbox{20cm}{"OAMHolo[$\psi$,d,6]", "BS[XXX,c,d]",  \\ "BS[XXX,a,c]", "LI[XXX,b,f]"} & $\ket{-1}$ & \pbox{20cm}{-FF2[-1]*FF3[-2]*FF4[-8] - FF2[-1]*FF3[-1]*FF4[-7] - \\ - FF2[-1]*FF3[0]*FF4[-6] + FF2[1]*FF3[-1]*FF4[-5] - \\ - FF2[-1]*FF3[2]*FF4[-4] + FF2[-1]*FF3[-8]*FF4[-2] + \\ + FF2[-1]*FF3[-7]*FF4[-1] + FF2[-1]*FF3[-6]*FF4[0] + \\ + FF2[-1]*FF3[-5]*FF4[1] + FF2[-1]*FF3[-4]*FF4[2]} \\ \hline
    (10,10,2) & \pbox{20cm}{"OAMHolo[$\psi$,d,-9]", "LI[XXX,a,f]",  \\ "BS[XXX,c,d]", "OAMHolo[XXX,a,2]", \\  "BS[XXX,a,d]"} & $\ket{-1}$ & \pbox{20cm}{-FF2[1]*FF3[-2]*FF4[-11] - FF2[1]*FF3[-1]*FF4[-10] - \\ - FF2[1]*FF3[0]*FF4[-9] - FF2[1]*FF3[1]*FF4[-8] - \\ - FF2[1]*FF3[2]*FF4[-7] + FF2[1]*FF3[-11]*FF4[-2] + \\ + FF2[1]*FF3[-10]*FF4[-1] + FF2[1]*FF3[-9]*FF4[0] + \\ + FF2[1]*FF3[-7]*FF4[2] - FF2[-1]*FF3[-8]*FF4[3]} \\ \hline
    (11,7,5) & \pbox{20cm}{"OAMHolo[$\psi$,b,6]", "OAMHoloSP2[XXX,d,-5]",  \\ "BS[XXX,a,b]", "LI[XXX,a,d]"} & $\ket{-1}+\ket{0}$ & \pbox{20cm}{-FF2[-2]*FF3[0]*FF4[-8] + FF2[-5]*FF3[2]*FF4[-7] - \\ - FF2[0]*FF3[0]*FF4[-6] + FF2[-5]*FF3[0]*FF4[-5] - \\ - FF2[2]*FF3[0]*FF4[-4] + FF2[-5]*FF3[-2]*FF4[-3] + \\ + FF2[-8]*FF3[0]*FF4[-2] + FF2[-5]*FF3[1]*FF4[-1] + \\ + FF2[-6]*FF3[0]*FF4[0] + FF2[-5]*FF3[-1]*FF4[1] + \\ + FF2[-4]*FF3[0]*FF4[2]} \\ \hline
    \textbf{DC=3} &   &   & \\ \hline
    (7,4,4) & \pbox{20cm}{"BS[$\psi$,b,e]", "LI[XXX,b,d]", \\ "BS[XXX,d,e]"} & $\ket{3}$ & \pbox{20cm}{FF2[3]*FF3[-3]*FF4[-3] - I*FF2[-2]*FF3[-2]*FF4[-3] - \\ - I*FF2[0]*FF3[0]*FF4[-3] - I*FF2[2]*FF3[2]*FF4[-3] + \\ + FF2[3]*FF3[-1]*FF4[-1] + FF2[3]*FF3[1]*FF4[1] + \\ + FF2[3]*FF3[3]*FF4[3]} \\ \hline
    (10,7,4) & \pbox{20cm}{"OAMHolo[$\psi$,d,6]", "BS[XXX,a,d]", \\ "LI[XXX,c,d]"} & $\ket{1}+\ket{8}$ & \pbox{20cm}{FF2[1]*FF3[-3]*FF4[-9] + FF2[1]*FF3[-1]*FF4[-7] + \\ + FF2[1]*FF3[1]*FF4[-5] + FF2[1]*FF3[3]*FF4[-3] + \\ + FF2[1]*FF3[4]*FF4[-2] + FF2[1]*FF3[6]*FF4[0] - \\ - FF2[-2]*FF3[-2]*FF4[2] - FF2[0]*FF3[0]*FF4[2] - \\ - FF2[2]*FF3[2]*FF4[2] + FF2[1]*FF3[8]*FF4[2]} \\ \hline
    (11,9,4) & \pbox{20cm}{"LI[$\psi$,a,d]", "LI[XXX,d,f]", \\ "BS[XXX,e,f]", "OAMHolo[XXX,c,5]", \\ "BS[XXX,c,d]", "OAMHolo[XXX,c,-4]", \\ "BS[XXX,c,e]"} & $\ket{-3}+\ket{0}$ & \pbox{20cm}{-I*FF2[3]*FF3[6]*FF4[-3] - I*FF2[3]*FF3[8]*FF4[-1] - \\ - I*FF2[3]*FF3[10]*FF4[1] + I*FF2[3]*FF3[1]*FF4[2] - \\ - I*FF2[3]*FF3[12]*FF4[3] + I*FF2[3]*FF3[3]*FF4[4] + \\ + I*FF2[2]*FF3[-2]*FF4[5] + I*FF2[0]*FF3[0]*FF4[5] + \\ + I*FF2[-2]*FF3[2]*FF4[5] + I*FF2[3]*FF3[5]*FF4[6] + \\ + I*FF2[3]*FF3[7]*FF4[8]} \\ \hline
    (12,7,7) & \pbox{20cm}{"OAMHolo[$\psi$,a,-6]", "BS[XXX,a,d]"} & $\ket{3}$ & \pbox{20cm}{FF2[3]*FF3[-3]*FF4[-9] + FF2[2]*FF3[-3]*FF4[-8] + \\ + FF2[1]*FF3[-3]*FF4[-7] + FF2[0]*FF3[-3]*FF4[-6] + \\ + FF2[-1]*FF3[-3]*FF4[-5] + FF2[-2]*FF3[-3]*FF4[-4] - \\ - FF2[-3]*FF3[-2]*FF4[-2] - FF2[-3]*FF3[-1]*FF4[-1] - \\ - FF2[-3]*FF3[0]*FF4[0] - FF2[-3]*FF3[1]*FF4[1] - \\ - FF2[-3]*FF3[2]*FF4[2] - FF2[-3]*FF3[3]*FF4[3]} \\ \hline    
  \end{tabular}
\end{center}
}}                                                                                      
  List 1: Experimental configuration for all found high-dimensional 3-partite entangled states. First column shows the Schmidt-Rank Vector. The second column shows the experimental configuration. There, the first element in the list is the first in the experiment (for example, the (2,2,2)-state's experiment can be calculated as LI[OAMHolo[$\psi$,c,-1],a,c]). The third line is the trigger in path A for the three-photon state in path B, C and D. The last gives the quantum state, where FF1, FF2 and FF3 stand for the photon in path B, C and D respectively.

\subsection{S5) Example: 3-dimensional GHZ-state (SRV=(3,3,3))}
The setup for the 3-dimensional GHZ-state in fig.(\ref{fig:multipartiteStates}A) consists of an OAM-Parity sorter, a mirror, a +2 Hologram and a beam-splitter. (Note that for simplicity, we do not write normalization-constants)

After down-conversion (neglecting the double-emissions from one crystal as they will be filtered in the four-fold coincidence detection in the end), we can write 

\begin{eqnarray}
\ket{\psi}&=&\left(\ket{0,0}+\ket{1,-1}+\ket{-1,1} \right)^2 =\nonumber\\
&=&\ket{0,0,0,0} + \ket{0,0,1,-1} + \ket{0,0,-1,1}  +\nonumber\\
&+&\ket{1,-1,0,0} + \ket{1,-1,1,-1} + \ket{1,-1,-1,1}  +\nonumber\\
&+&\ket{-1,1,0,0} + \ket{-1,1,1,-1} + \ket{-1,1,-1,1} 
\label{eq:GHZ1}
\end{eqnarray}
After the OAM-Parity sorter (again, neglecting double-photons in one detector-arm, because they will be filtered by four-fold coincidence detection). From here on, only events from the two crystals with the same parity survive:

\begin{eqnarray}
\ket{\psi}=\ket{0,0,0,0}+\ket{1,-1,1,-1}+\ket{1,-1,-1,1}+\ket{-1,1,1,-1}+\ket{-1,1,-1,1}
\label{eq:GHZ2}
\end{eqnarray}
For a Trigger in A with $\left(\ket{0}+\ket{1}\right)$, the state has a SRV of (3,3,2), as one can see in the list above. Photon C and D reside in 3 dimensions while photon B lives in a 2-dimensional space. That is because photon B is perfectly anti-correlated with photon A, which is the 2-dimensional trigger. We want to increase the dimension of photon B to 3. The general idea is remove the perfect anti-correlation by mixing the trigger with photon C. At this stage, a BS between A and C would lead to Hong-Ou-Mandel interference between the 1st, 3rd and 4th term, which effectively removes those terms from the state and does not lead to a 3-dimensional GHZ state. In order to prevent this from happening, the photon in A is shifted by -2. This prevents HOM interference between those three terms, and removes the 2nd term instead. One additional subtle but significant trick is the usage of the mirror in order to be not vulnerable to higher-order terms (without the mirror, the state would become a 2-dimensional GHZ if higher-order modes in SPDC are considered).
The mirror in arm A leads to: 
\begin{eqnarray}
\ket{\psi}=\ket{0,0,0,0}+\ket{-1,-1,1,-1}+\ket{-1,-1,-1,1}+\ket{1,1,1,-1}+\ket{1,1,-1,1}
\label{eq:GHZ3}
\end{eqnarray}
And the hologram of -2 in A transforms the state to
\begin{eqnarray}
\ket{\psi}=\ket{-2,0,0,0}+\ket{-3,-1,1,-1}+\ket{-3,-1,-1,1}+\ket{-1,1,1,-1}+\ket{-1,1,-1,1}
\label{eq:GHZ4}
\end{eqnarray}
In the next step, a beam-splitter will be placed between arm A and C.
\begin{eqnarray}
\ket{\psi}&=&\ket{0,0,-2,0}-\ket{2,0,0,0}+\ket{1,-1,-3,-1}-\ket{3,-1,-1,-1}+\ket{-1,-1,-3,1}-\ket{3,-1,1,1} +\nonumber\\
&\mathbin{\color{red}+}&\mathbin{\color{red}\ket{1,1,-1,-1}-\ket{1,1,-1,-1}}+\ket{-1,1,-1,1}-\ket{1,1,1,1}
\label{eq:GHZ5}
\end{eqnarray}
The red terms cancel because of destructive interference. This is due to the Hong-Ou-Mandel effect, which occurs if the OAM of two incoming photons from two different arms in a beam splitter are opposite, which leaves us with the state
\begin{eqnarray}
\ket{\psi}&=&\ket{0,0,-2,0}-\ket{2,0,0,0}+\ket{1,-1,-3,-1}-\ket{3,-1,-1,-1}+\ket{-1,-1,-3,1}-\ket{3,-1,1,1} +\nonumber\\
&+&\ket{-1,1,-1,1}-\ket{1,1,1,1}
\label{eq:GHZ6}
\end{eqnarray}
If we now use the photon A as Trigger for $\left(\ket{0}+\ket{1}\right)$, the photons in B, C and D will be in the state:
\begin{eqnarray}
\ket{\psi}&=\ket{0,-2,0}+\ket{-1,-3,-1}-\ket{1,1,1}
\label{eq:GHZ6}
\end{eqnarray}
This state fulfills the criterion for a high-dimensional GHZ-state as stated above, and with local unitaries can be transformed into $\ket{\psi}=\ket{0,0,0}+\ket{1,1,1}+\ket{2,2,2}$.

\subsection{6) Cyclic rotations in a high-dimensional space}
Here we list cyclic rotation found by the algorithm, either with OAM only, or with OAM and polarization, or with OAM and polarization and path. In all examples, $\psi$ stands for the initial state and $XXX$ stand for the state after the previous element.

\subsubsection{4-cyclic OAM rotation}
Experimental configuration:
\begin{eqnarray}
\TN{BS}[\psi,\TN{a},\TN{b}] &\to& \TN{DP}[XXX,\TN{b},1] \to \TN{Reflection}[XXX,\TN{b}] \to \TN{BS}[XXX,\TN{a},\TN{b}]\nonumber\\
&\to& \TN{Reflection}[XXX,\TN{a}] \to \TN{BS}[XXX,\TN{a},\TN{b}] \to \TN{DP}[XXX,\TN{b},1]\nonumber\\
&\to& \TN{Reflection}[XXX,\TN{b}] \to \TN{BS}[XXX,\TN{a},\TN{b}] \to \TN{OAMHolo}[XXX,\TN{a},1]\nonumber\\         
\label{eq:4cyclic}
\end{eqnarray}

Operation:
\begin{eqnarray}
\ket{-1} \to \ket{0} \to \ket{1} \to \ket{2} \to \ket{-1}      
\label{eq:4cyclicOp}
\end{eqnarray}
The number in the ket stands for the OAM. This experiment has been performed in our laboratories\cite{schlederer2015cyclic}.

\subsubsection{3-cyclic OAM+Polarisation rotation}
Experimental configuration:
\begin{eqnarray}
\TN{HWP}[\psi,\TN{a}] &\to& \TN{Reflection}[XXX,\TN{a}] \to \TN{PBS}[XXX,\TN{a},c] \to \TN{OAMHolo}[XXX,\TN{a},2]\nonumber\\
&\to& \TN{Reflection}[XXX,\TN{a}] \to \TN{PBS}[XXX,\TN{a},c] \to \TN{BS}[XXX,\TN{a},\TN{b}]\nonumber\\
&\to& \TN{DP}[XXX,\TN{b},2] \to \TN{Reflection}[XXX,\TN{b}] \to \TN{BS}[XXX,\TN{a},\TN{b}]\nonumber\\
&\to& \TN{HWP}[XXX,\TN{b}] \to \TN{BS}[XXX,\TN{a},\TN{b}] \to \TN{DP}[XXX,\TN{b},2]\nonumber\\
&\to& \TN{Reflection}[XXX,\TN{b}] \to \TN{BS}[XXX,\TN{a},\TN{b}]
\label{eq:3cyclic}
\end{eqnarray}

Operation:
\begin{eqnarray}
\ket{-2,V} \to \ket{-2,H} \to \ket{0,V} \to \ket{-2,V}
\label{eq:3cyclicOp}
\end{eqnarray}
The number in the ket stands for the OAM, H and V stand for horizontal and vertical polarisation.

\subsubsection{6-cyclic OAM+Polarisation rotation}
Experimental configuration:
\begin{eqnarray}
\TN{HWP}[\psi,\TN{a}] &\to& \TN{Reflection}[XXX,\TN{a}] \to \TN{PBS}[XXX,\TN{a},\TN{c}] \to \TN{OAMHolo}[XXX,\TN{a},2]\nonumber\\
&\to& \TN{Reflection}[XXX,\TN{a}] \to \TN{PBS}[XXX,\TN{a},\TN{c}] \to \TN{BS}[XXX,\TN{a},\TN{b}]\nonumber\\
&\to& \TN{DP}[XXX,\TN{b},2] \to \TN{Reflection}[XXX,\TN{b}] \to \TN{BS}[XXX,\TN{a},\TN{b}]\nonumber\\
&\to& \TN{HWP}[XXX,\TN{b}] \to \TN{BS}[XXX,\TN{a},\TN{b}] \to \TN{DP}[XXX,\TN{b},2]\nonumber\\
&\to& \TN{Reflection}[XXX,\TN{b}] \to \TN{BS}[XXX,\TN{a},\TN{b}]
\label{eq:6cyclic}
\end{eqnarray}

Operation:
\begin{eqnarray}
\ket{-4,H} \to \ket{-2,H} \to \ket{0,V} \to \ket{2,H} \to \ket{4,V} \to \ket{-4,H}
\label{eq:6cyclicOp}
\end{eqnarray}
The number in the ket stands for the OAM, H and V stand for horizontal and vertical polarisation.

\subsubsection{8-cyclic OAM+Polarisation rotation}
Experimental configuration:
\begin{eqnarray}
\TN{PBS}[\psi,\TN{a},\TN{b}] &\to& \TN{BS}[XXX,\TN{b},\TN{c}] \to \TN{DP}[XXX,\TN{c},1] \to \TN{Reflection}[XXX,\TN{c}]\nonumber\\
&\to& \TN{BS}[XXX,\TN{b},\TN{c}] \to \TN{Reflection}[XXX,\TN{b}] \to \TN{BS}[XXX,\TN{b},\TN{c}]\nonumber\\
&\to& \TN{DP}[XXX,\TN{c},1] \to \TN{Reflection}[XXX,\TN{c}] \to \TN{BS}[XXX,\TN{b},\TN{c}]\nonumber\\
&\to& \TN{OAMHolo}[XXX,\TN{b},1] \to \TN{PBS}[XXX,\TN{a},\TN{b}] \to \TN{HWP}[XXX,\TN{a}]
\label{eq:8cyclic}
\end{eqnarray}

Operation:
\begin{eqnarray}
\ket{-1,V} \to \ket{-1,H} \to \ket{0,V} \to ... \to \ket{2,H} \to \ket{-1,V}
\label{eq:8cyclicOp}
\end{eqnarray}
The number in the ket stands for the OAM, H and V stand for horizontal and vertical polarisation.

\subsubsection{14-cyclic OAM+Polarisation+Path rotation}
Experimental configuration:
\begin{eqnarray}
\TN{Reflection}[\psi,\TN{a}] &\to& \TN{OAMHolo}[XXX,\TN{a},2] \to \TN{Reflection}[XXX,\TN{a}] \to \TN{OAMHolo}[XXX,\TN{a},-2]\nonumber\\
&\to& \TN{PBS}[XXX,\TN{a},\TN{b}] \to \TN{HWP}[XXX,\TN{a}] \to \TN{PBS}[XXX,\TN{a},\TN{b}]\nonumber\\
&\to& \TN{Reflection}[XXX,\TN{b}] \to \TN{OAMHolo}[XXX,\TN{a},2] \to \TN{Reflection}[XXX,\TN{a}]\nonumber\\
&\to& \TN{BS}[XXX,\TN{a},\TN{b}] \to \TN{DP}[XXX,\TN{b},2] \to \TN{Reflection}[XXX,\TN{b}]\nonumber\\
\to \TN{BS}[XXX,\TN{a},\TN{b}]
\label{eq:14cyclic}
\end{eqnarray}

Operation:
\begin{eqnarray}
\ket{0,H,a} &\to& \ket{-2,H,b} \to \ket{-4,H,b} \to \ket{-8,H,b} \to \ket{10,V,b}\nonumber\\
&\to& \ket{-6,H,a} \to \ket{8,H,a} \to \ket{6,H,b} \to \ket{4,H,a}\nonumber\\
&\to& \ket{0,H,b} \to \ket{2,V,b} \to \ket{2,V,a} \to \ket{2,H,a}\nonumber\\
&\to& \ket{0,H,a}
\label{eq:14cyclicOp}
\end{eqnarray}
The number in the ket stands for the OAM, H and V stand for horizontal and vertical polarization, and a and b stand for the two different possible paths.

\subsection{S7) Learning algorithm}
In the second example involving cyclic operations, the algorithm extends its own set of basic elements autonomously, based on the properties of the longest cycle. It saves elements that have large cycles and experiments with non-trivial coupling between different degrees-of-freedom. Additionally, elements already learned can be "forgotten" to improve variability and prevent dead-ends, as some of them might even have negative effects on the probability of finding new experiments. The decision of which elements are forgotten at which times is purposefully random. Even though it would be possible to weight the elements for past usefulness, it would introduce a bias on similar solutions that we wanted to prevent.

\subsection{S8) Simplification of experiments}
After an experiment is found, it is simplified. For that, three different methods are used. The first one removes elements from the experiment and calculates whether the resulting state or transformation is still performed the same way. Such simplifications could remove elements in paths that are not accessed. An example, in which it is necessary to remove multi elements at the same time, is the following: Four beam splitters after each other form two Mach-Zehnder interferometers. Those have no effect if the phases are set correctly, but can only be removed together.

In the second method, it is tried to replace more complicated elements (such as LI, PBS or DP) by mirrors. This works in cases where the only specific modes access the element (for instance, if only vertically polarized photons access a PBS).

A third method tries to simplify the path structure of the experiment by rearranging the paths. For example, if two PBSs are used after each other, one output of the second PBS will never be used, thus the second PBS can be removed and one path can be removed completely. 

Those three methods are applied iteratively, until no simplification is possible anymore.